\documentstyle{article}
\newenvironment{namelist}[1]{%
\begin{list}{}
    {
     
     \settowidth{\labelwidth}{#1}
     \setlength{\leftmargin}{1.1\labelwidth}
    }
  }{%
\end{list}}
\begin{document}
\newcommand{\Bog}{Bogot\'{a}}
\raggedbottom

\begin{center}

{TOPOLOGICAL ANALYSIS OF COBE-DMR CMB MAPS} \\

\vspace{4mm}

Sergio Torres \footnote{On leave from Universidad de los Andes,
and Centro Internacional de F\'{i}sica, Bogot\'{a},
Colombia.
E-mail: 40174::torres, torres@celest.lbl.gov}  \\

\vspace{4mm}
International Center for Relativistic Astrophysics,
Univ. di Roma 1, Ple Aldo Moro, 5 - 00185, Rome, Italy
\end{center}

\vspace{4mm}

{\it Subject headings:} cosmic microwave background

\vspace{4mm}


{\bf Abstract.} Geometric characteristics of random fields
are exploited to test the consistency of density perturbation
model spectra with COBE data.  These CMB maps are analyzed
using the number of anisotropy hot spots and their boundary curvature.
CMB maps which account for instrumental effects and sky coverage
are Monte Carlo generated.  These simulations show that a scale invariant
Harrison-Zeldovich primordial Gaussian density fluctuation spectrum
is consistent with the data.  The CMB fluctuation coherence angle,
based on boundary curvature, gives a spectral index $n = 1.2 \pm 0.3$.

\section{Introduction}

Gravitational collapse models of structure formation, predict
that at large angular scales the cosmic microwave background (CMB)
will exhibit random fluctuations. The gravitational potential on
the surface which last scatter CMB photons (Sachs \& Wolfe 1967) is
responsible for these fluctuations.  Thus, the primordial matter density
spectrum determines the amplitudes of these fluctuations which in
$k-$space is $P(k) = \left| \delta_k \right|^2$ and can be represented
by a power law, $P(k) \propto k^n$, with spectral index $n$.  The
inflationary model requires a scale invariant spectrum with $n = 1$
(Harrison 1970; Zeldovich 1972).

$P(k)$ can be inferred from the CMB temperature autocorrelation
function (acf).  Alternatively, $P(k)$ models can be tested statistically
using the hot spots in CMB maps.  The most important advantage of
this second approach is that, unlike acf computations, topological
analysis does not assume an ergodic radiation field (Cay\'{o}n et al. 1991;
Scaramella and Vittorio 1990), and so
provides normalization independent power spectrum information.  This
later analysis can also be used to diagonse non-random CMB map components
(e.g. foreground signals); as an unbias coherence angle estimator of an
underlying random field (Adler 1981); and to verify the Gaussian
nature of primordial fluctuations and thus test structure formation
models relying on primordial seeds (e.g. cosmic strings
(Bennett et al. 1992) or texture (Park et al. 1991)).
Also the existence of non-Gaussian sky temperature contributions
from non-ergodic behavior can be determined using Monte Carlo simulations.

The first year COBE-DMR maps (Smoot et al. and references therein) are
topologically analyzed to check their consistency with the
Harrison-Zeldovich power spectrum which a preliminary hot spot analysis
found to be the case (Gurzadyan and Torres 1993).

\section{The geometry of random fields}

The random field excursion set is the domain of all points in which the
field takes on values $T \ge T_{\nu} =  \nu \sigma$, where $\sigma$
and $\nu$ are the temperature field standard deviation and threshold,
respectively.  Topological descriptors which have been used to
characterize CMB excursion regions (Sazhin 1985; Vittorio and
Juszkiewics 1987; Bond and Efstathiou 1987; Coles and Borrow 1987;
Coles 1988a; Coles 1988b; Gott et al. 1990;
Mart\'{i}nez-Gonz\'{a}lez 1989) are:  The number of spots with temperature
above $T_{\nu}$, and below $-T_{\nu}$, $N_{\nu}$;  the mean area of these
spots, $A_{\nu}$;  the total spot boundary curvature,
or genus $G_{\nu}$;  the contour level lengths, $L_{\nu}$;  the
total spot excursion area, $a_{\nu}$;  the number of up crossings
(one dimension) and the Euler-Poincar\'{e} characteristic (two dimensions
(Adler 1981)).

Excursion set geometry theory allows some descriptors to be derived
analytically. (e.g. For two-dimensional, stationary, isotropic, random,
Gaussian fields the geometric properties depend only on the acf and its
second derivative at zero lag.)  A more compact representation is the
coherence angle,
$\theta_c^2 = - C(0)/C^{\prime\prime}(0).$
The mean number of spots on the $4\pi$ surface, the mean genus,
and the total excursion set area are:
\begin{equation}
\langle N_{\nu} \rangle =
\frac{2}{\pi \theta_c^2} \frac{\exp(\nu^2)}{\mbox{erfc}(\nu/\sqrt2)}  ,
\end{equation}
\begin{equation}
\langle G_{\nu} \rangle =
\left( \frac{2}{\pi}  \right)^{1/2}
\frac{\nu}{\theta_c^2}\exp(-\frac{\nu^2}{2})   , \;  \mbox{and}
\end{equation}
\begin{equation}
\langle a_{\nu} \rangle =
2\pi  \mbox{erfc}(\nu/\sqrt2).
\end{equation}
The total excursion set area is independent of  $\theta_c$ and so can be used
to identify systematic effects in the data.

\section{Computing topological descriptors}

COBE data is apportioned to pixels by projecting the celestial sphere onto
a circumscribed sky cube, each cube side being divided into 1024 pixels
(Torres et al. 1989).
The resulting 6144 pixels, each  have $\approx 2.9^{\circ} \times
2.9^{\circ}$ angular dimensions comparable to COBE-DMR angular resolution.
The pixel face and Cartesian coordinate identifier allow efficient algorithms
to be constructed so that the data can be accessed in a form close to that of
the original.

Since hot spots are defined in terms of a threshold, $\nu$, a binary map is
constructed for each threshold level, a 1 appearing in pixels whose
temperature is greater than or equal to $T_{\nu}$.  Hot spots are located on
these bit maps by forming tree data structures.  When traversing the sky
cube the first 1 pixel encountered becomes the root node and all contiguous
1 pixels become nodes of that tree.  A tree is completed when all neighboring
pixels are 0.  The hot spot area is the number of nodes in the tree.  The
contour level length is found by counting the neighboring 1 pixels of each
node.
The total curvature index, or genus, of a bit map is the number of
hot spots minus the number of holes (Gott et al. 1992) which are those spots
found using the hot spot algorithm recursively on the negative of the bit
map minus one.

Simulated Gaussian random field maps are generated and their hot spot number
densities and genus, areas, and contour lengths evaluated.  Figure 1 shows
the simulated total area $a$ mean values with theoretical $a_{\nu}$ values
while Fig. 2 does the same for genus.  The agreement is good.

Excluding the equatorial band to remove contributions from our galaxy
introduces a boundary thus enhancing $G_{\nu}$ and $N_{\nu}$ at low
thresholds.  To account for this artifact all Monte Carlo generated data
sets are treated identically to that of the experimental data.

\section{Monte Carlo simulations}

A COBE-DMR CMB map can be simulated by assigning to each pixel
a temperature given by an expansion of real spherical harmonics
as defined in Smoot et al. (1991) weighted by the beam filter
coefficients $W_{\ell}$.
The finite beam width, $\sigma_{\mbox{b}}$, of the instrument
enters in $W_{\ell}$ as a high-pass filter
(Vittorio, Scaramela 1988):
\begin{equation}
W_{\ell} =
\exp \left[-\frac{1}{2} \sigma_{\mbox{b}}^2 \ell (\ell + 1).
\right]
\end{equation}
For random Gaussian fields, the harmonic coefficients of the expansion
are random variables with zero mean, are Gaussianlly distributed, and
have variance determined by the primordial matter density perturbation
spectrum.

On the scale of the DMR beam $(> 2^{\circ})$ only fluctuations of the
gravitational potential on the surface of last scattering affect the CMB
isotropy and so the Sachs-Wolfe effect alone suffices to determine the
variances.  For power law spectra, $P(k) \propto k^n$, the Bond-Efstathiou
formula (1987) for the variances is adequate since only contributions
for $\ell$ up to $\sim 30$ are important after beam width filtering and
further Gaussian smoothing.  Instrumental noise and sky coverage are
accounted for by adding to each pixel noise equals to a Gaussian random
number with dispersion equal to the noise level per measurement divided by
$\sqrt{N_{obs}}$, where the number of observations is included in
the original sky maps.  To estimate noise for a single measurement
the observed temperature squared is apportioned among bins of equal width
in $1/N_{obs}$, and the average in each bin is used. The slope of the
$\langle T_o^2 \rangle$ vs. $1/N$ plot gives the noise level squared.
The levels for each channel at each of the three DMR frequencies were
within 7\% of nominal values.

Two maps are Monte Carlo generated, each with its corresponding noise level,
from which sum and difference maps are constructed and $2.9^{\circ}$ Gaussian
smoothed.  Each produced sky map is one realization of the ensamble
from which $G_{\nu}$, $N_{\nu}$, $A_{\nu}$, and $L_{\nu}$ are extracted.
After 400 realizations the respective means and dispersions are computed.

Figure 3 shows the means and their 1 sigma error bars for the number
of hot spots  as seen by the 53 GHz DMR radiometer observing
a universe with primordial Harrison-Zeldovich fluctuations.  Also shown are
the spot numbers  obtained from maps made with pure instrumental
noise.  The separation between these two curves indicates both the power of a
statistical test and the discrimination efficiency of a topological
descriptor.  Noisy maps have noise and signal curves in close proximity.
$L_{\nu}$ and $A_{\nu}$ have such close noise and signal curves and so can not
discriminate between power spectra with indices near one.  Genus and
hot spot number are the only descriptors used to test hypothesis.

To compare COBE-DMR data with cosmological models $\chi^2$ and the cumulative
probability distribution of $\chi^2$ are constructed from Monte Carlo data
(Gott et al. 1992).  For each Monte Carlo map the topological descriptor
values $U_{\nu}$ evaluated at thresholds $\nu: 1...25$ are used to compute,
\begin{equation}
\chi^2 =
\frac{1}{24}\sum_{j=1}^{25} \frac{(U_j - \langle U_j \rangle)^2}{\sigma_j^2},
\end{equation}
where $\langle U_j \rangle$ and $\sigma_j$ are calculated from the Monte
Carlo simulations for that cosmological model. Table 1 shows the $\chi^2$
and probabilities for the 53 and 90 sum maps and their weighted average.

\section{Preparation of data}

In this analysis four of the six DMR sky maps -- 53-A, -B, 90-A, and -B --
are used.  The data is in the sky cube format using Geocentric Ecliptic
coordinates.  Before the topological analysis the dipole and $2 \mu$K
kinematic quadrupole components are removed, the maps are rotated to galactic
equatorial coordinates, and to reduce noise the maps have an additional
Gaussian smoothing of $2.9^{\circ}$.
An equatorial band of $|b|<30^{\circ}$
is excluded from the analysis.
Cosmological hot and cold spots
will exists in all four maps as unresolved sources, so a weighted average of
the four maps is made to reduce the noise.

As a data integrity and analysis software check the diploe, rms $10^{\circ}$
sky variation, and quadrupole are computed and compared with COBE results.
The dipole and quadrupole are obtained by a least
squares fit of the harmonic expansion to the map and are in agreement
with Smoot et al. (1992).

\section{Analysis and Conclusions}

Figure 1, a plot of $a_{\nu}$ versus $\nu$, illustrates what would be
obtained for a pure random field independent of its coherence angle.
In the same figure,
$a$ versus threshold for the 31 GHz DMR sum map, shows large
deviations which increase with decreasing galactic cut indicating galactic
contamination at high latitudes.  For the 53 and 90 sum maps no deviations
beyond 1 sigma of Fig. 1 are seen above 25 degrees.

Genus and number of spots were used to test for the consistency of a
primordial Harrison-Zeldovich perturbation spectrum. The null hypothesis
that the structure in COBE maps is due solely to instrumental noise is rejected
with high confidence as seen in Table 1 and Fig. 3.  The contribution from
known infrared, radio, and x-ray point sources is negligible (Bennett et
al. 1993) so the observed structure must be of cosmological origin.

Genus and number of spots computed on Monte Carlo realizations of CMB maps
with a Harrison-Zeldovich power spectrum normalized to $16 \mu$K Q$_{rms}$
is found to be consistent with DMR data to a high confidence, as seen in
Table 1, as are other spectral indices.  To see the genus
dependence on the spectral index, several sets of Monte Carlo realizations
are generated with fixed Q$_{rms}$ normalization and variable $n$.  For each
index the genus curve as a function of threshold $\nu$ is fit using (3) to
obtain the best $\theta_c$.  The coherence angle dependence on spectral index
is almost linear, as seen in Fig. 4.  With $\theta_c$ errors obtained from
Monte Carlo simulations the best value of $n$ is found by fitting the COBE
genus curves to determine their coherence angle which for the 53 GHz sum DMR
map is  $4.5^{\circ} \pm 0.1^{\circ}$ which implies a $1.2 \pm 0.3$ spectral
index in agreement with Smoot et al. (1992) using a acf fit.

{\it Acknowledgements:} I thank W. Peter Trower for his suggestions
and help in editing the manuscript,
Prof. Remo Ruffini,
Laura Cay\'{o}n,
Roberto Fabbri,
Al Kogut,
Enrique Mart\'{i}nez-Gonz\'{a}lez, and
George Smoot for many useful comments and conversations.
The support of the SuperCompter
Computations
Research Institute of Florida State University, where some
of the Monte Carlos were done,
is greatly appretiated.
This work has been supported by
Colciencias of Colombia
project \# 1204-05-012-91, and the European Comunity under contract
No. CI1-CT92-0013.  The COBE datasets were developed by the NASA
Goddard Space Flight Center under the guidance of the COBE Science
Working Group and were provided by the NSSDC.

\newpage

\vspace{5mm}
\vspace{2pt}
\raggedbottom
\begin{center}
TABLE 1 \\
$\chi^2$ and its probability for number of spots and genus of
COBE-DMR maps
\vspace{4mm}
\begin{tabular}{l@{\quad}l@{\quad}l@{\quad}l@{\quad}l}
\hline
Map &
$\chi^2_{\mbox{N}}$ & $P(\chi^2 \geq  \chi^2_{\mbox{N}})$ &
$\chi^2_{\mbox{G}}$ & $P(\chi^2 \geq  \chi^2_{\mbox{G}})$ \\
\hline
53     & 0.62  & 95.4     &  0.72  &  83.3   \\
90     & 1.48  & 13.9     &  1.73  &  8.4    \\
5390   & 1.03  & 47.1     &  1.03  &  47.1   \\
$53^*$ & 6.75  & $<1/400$   &  7.51  &  $<1/400$ \\
$90^*$ & 3.59  & $<1/400$   &  4.38  &  $<1/400$ \\
\hline
\end{tabular}
\vspace{4mm}
\end{center}

NOTE. - The first three rows are the $\chi^2$ statistic (5) for the
number of spots $\chi^2_{\mbox{N}}$ and genus $\chi^2_{\mbox{G}}$
and their probabilities (\%) when the Harrison-Zeldovich hypothesis is
assumed. The 5390 entry refers to the statistics of the weighted
averaged map of DMR 53 and 90 sum maps. Last two rows give the
same statistics when the null hypothesis is assumed in the Monte Carlo
procedure.

\newpage

{\bf \noindent References}

\begin{namelist}{xxx}
\setlength{\parskip}{0.0mm}
\setlength{\parsep}{0.0mm}
\setlength{\itemsep}{0.0mm}
\setlength{\labelsep}{0.0mm}
\raggedbottom

\item[Adler], R. J. 1981, The geometry of Random Fields (New York: Wiley)
\item[Gott], J. R., et al. 1990, ApJ, 352, 1
\item[Gurzadyan], V., and Torres, S. 1993, in The Present and
Future of the Cosmic Microwave Background, Edts. Mart\'{i}nez-Gonz\'{a}lez, E.
Cay\'on, L., and Saez, J. L., (in press)
\item[Bennett], C. et al. 1993, ApJ, 414, L77
\item[Bennett], D. P., Stebbins, A., \& Bouchet, F. R. 1992, ApJ, 399, L5
\item[Bond], J. R., and  Efstathiou, G. 1987, MNRAS, 226, 655
\item[Cay\'{o}n], L.,
Mart\'{i}nez-Gonz\'{a}lez, E., and
Sanz, J. L. 1991, MNRAS, 253, 599
\item[Coles], P. 1988a, MNRAS, 231, 125
\item[Coles], P. 1988b, MNRAS, 234, 509
\item[Coles], P. and Barrow, D. 1997, MNRAS, 228, 407
\item[Harrison], E. R. 1970, Phys. Rev. D, 1, 2726
\item[Mart\'{i}nez-Gonz\'{a}lez], E., and Sanz, J. L., 1989, MNRAS, 237, 939
\item[Park], C., Spergel, D. N., and Turok, N. 1991, ApJ, 372, L53
\item[Sachs], K., and Wolfe, A. M. 1967, ApJ, 147, 73
\item[Sazhin], M. V. 1985, MNRAS, 216, 25p
\item[Scaramella], R. and Vittorio, N. 1988, ApJ, 331, L53
\item[Smoot], G.F. et al. 1992, ApJ, 396, L1
\item[Smoot], G.F. et al. 1991, ApJ,  371, L1
\item[Torres], S., et al. 1989, in {\it Data Analysis in Astronomy}, ed.
V. di Gesu, L. Scarsi \& M. C. Maccarone (Erice, 1988), 40, pp. 319-333
\item[Scaramella], R., and Vittorio, N. 1990, ApJ, 353,372
\item[Vittorio], N. and Juszkiewicz, R. 1987, ApJ, 314, L29
\item[Zeldovich], Ya. B. 1972, MNRAS, 160, 1P
\end{namelist}

\newpage
\raggedbottom
\vspace{7mm}
\noindent

{\bf FIGURE CAPTIONS}

\vspace{5mm}

\begin{namelist}{xxxxxxx}
\setlength{\parskip}{0.0mm}
\setlength{\parsep}{0.0mm}
\setlength{\itemsep}{0.0mm}
\setlength{\labelsep}{0.0mm}
\raggedbottom

\item[Figure 1]. Mean value of the total area of excursion
divided by $2 \pi \mbox{erfc} (\nu/\sqrt2)$ at
different threshold levels.
The horizontal line correspond to the expected theoretical value for a
random field, independent of its coherence angle.
Open squares with 1 sigma error bars are for Monte Carlo generated maps.
Circles are for the 31(A+B) DMR maps with a $30^{\circ}$
(filled in circles) and a $10^{\circ}$ (open circles) galactic cut.
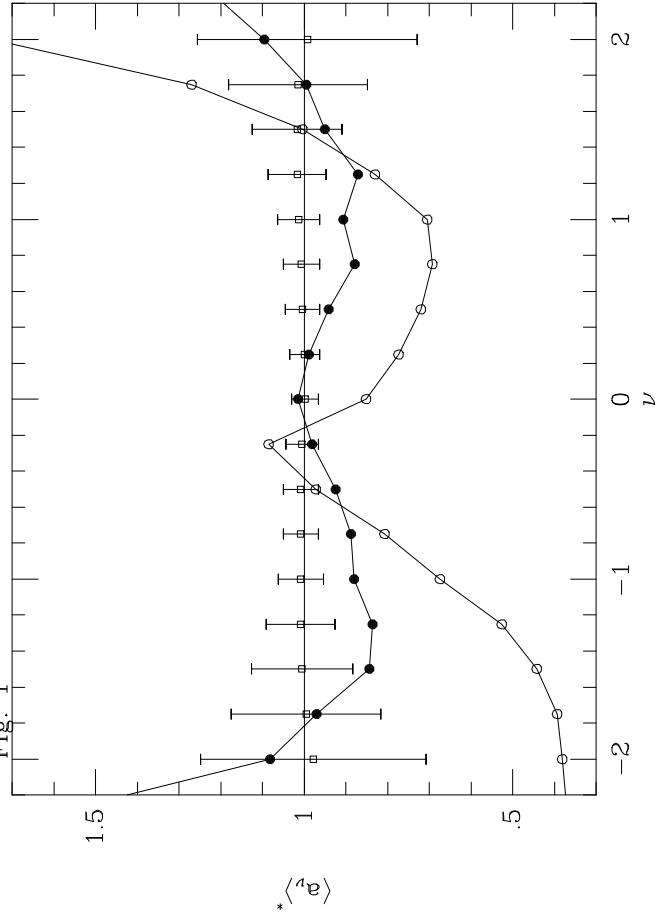
\item[Figure 2]. Mean values of the genus as a function of
threshold for Monte Carlo simulations of a random gaussian maps.
The solid curve is the analytical function.
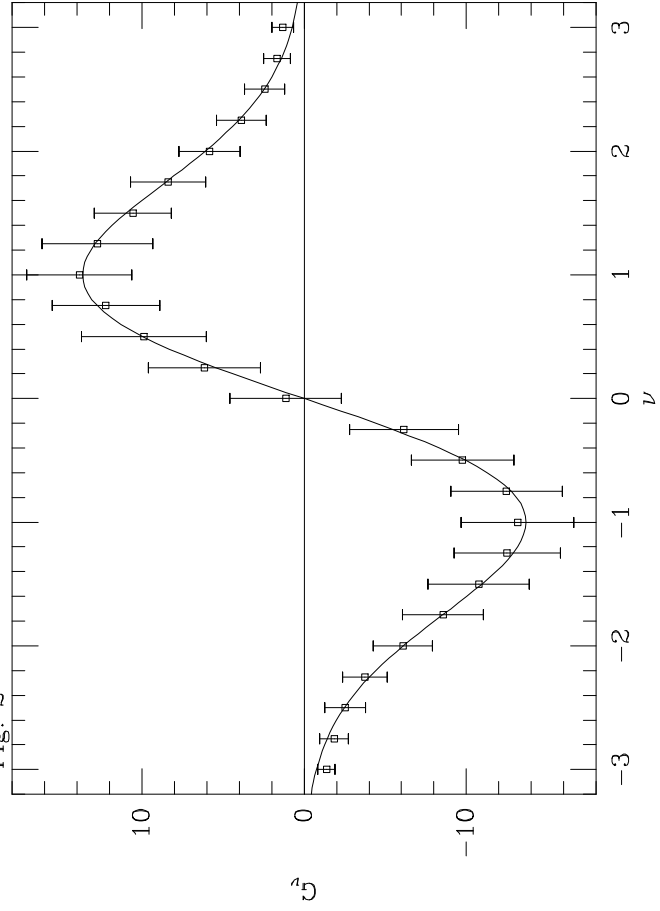
\item[Figure 3]. Number of spots normalized to the area of the sphere
including cold and hot spots.
Upper curve is for noise maps alone (open squares).
The other curve (open squares) is for
maps generated with noise and cosmological fluctuations with a
Harrison-Zeldovich spectrum (16 $\mu$K normalization).
Superposed is the data obtained from
the 53(A+B) DMR maps (filled in triangles).
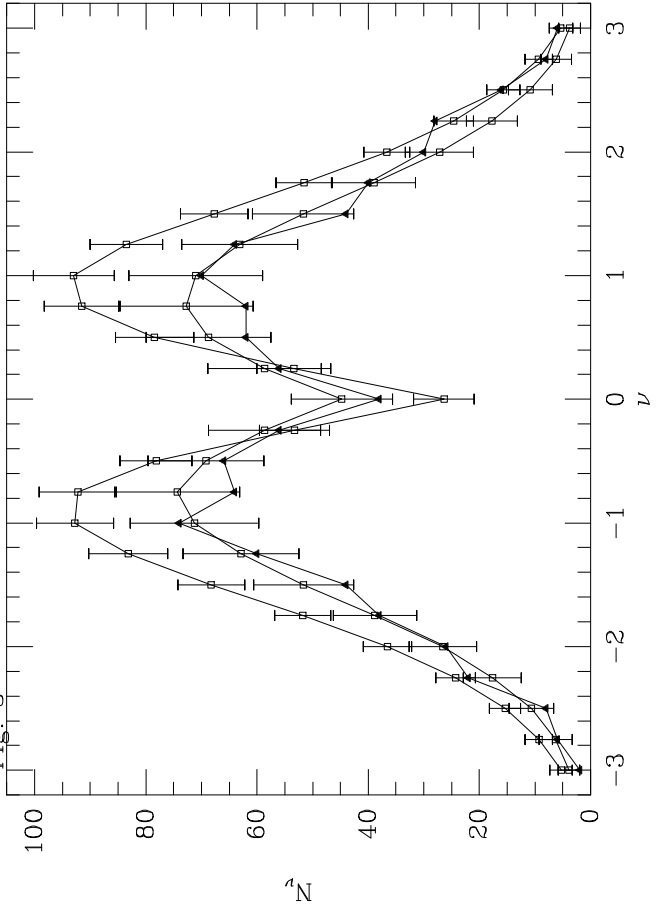
\item[Figure 4]. Coherence angle versus spectral index as obtained
from the Monte Carlo runs. Solid curve is a fit to a parabola.
\end{namelist}

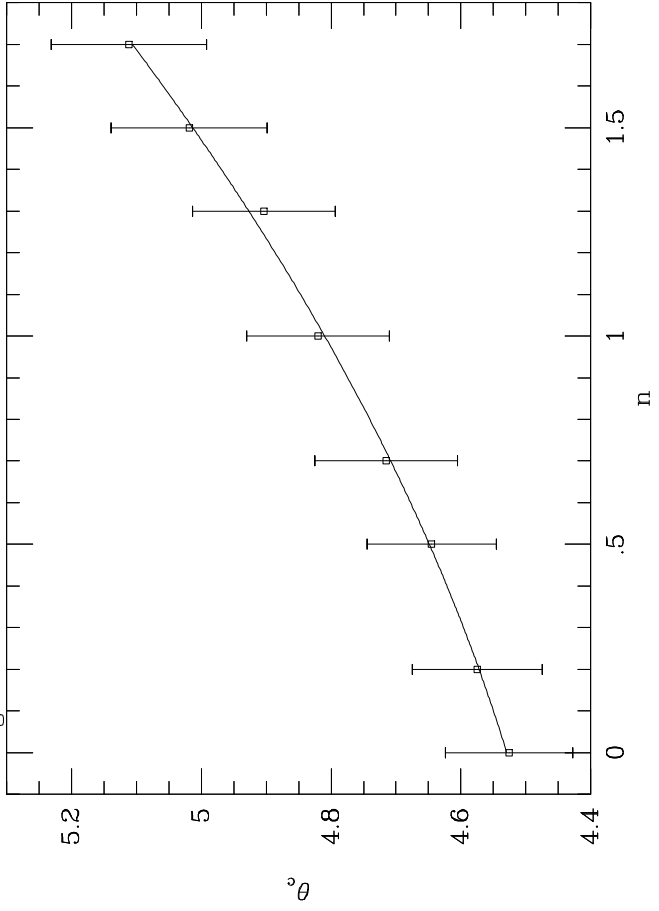
\end{document}